\documentclass[aps,prb, twocolumn,showpacs,preprintnumbers,superscriptaddress]{revtex4}

\usepackage{graphicx}
\usepackage{dcolumn}
\usepackage{bm}

\begin{document}


\title{Enhanced superconducting properties of double-chain based superconductor  Pr$_{2}$Ba$_{4}$Cu$_{7}$O$_{15-\delta}$ synthesized by citrate pyrolysis technique}

\author{Kazuma Honnami}
\author{Michiaki Matsukawa} 
\email{matsukawa@iwate-u.ac.jp }
\author{Tatsuya Senzaki}
\author{Tomoki Toyama}
\author{Haruka Taniguchi}
\author{Koichi Ui}
\affiliation{Faculty of Science and Engineering, Iwate University, Morioka 020-8551, Japan}

\author{Takahiko Sasaki}
\author{Kohki Takahashi}
\affiliation{Institute for Materials Research, Tohoku University, Sendai 980-8577, Japan}

\author{Makoto Hagiwara}
\affiliation{Kyoto Institute of Technology, Kyoto 606-8585, Japan}
\author{Fumihiko Ishikawa}
\affiliation{Department of Physics, Niigata University, Niigata 950-2181, Japan}

\date{\today}

\begin{abstract}
We report the enhanced superconducting properties of double-chain based superconductor  Pr$_{2}$Ba$_{4}$Cu$_{7}$O$_{15-\delta}$ synthesized by the  citrate pyrolysis technique.
 The reduction heat treatment in vacuum results in the appearance of superconducting state with $T_\mathrm{c}$=22-24 K, accompanied by the higher residual resistivity ratios.
The superconducting volume fractions are estimated from the ZFC data to be 50$\sim55\%$, indicating the bulk superconductivity. 
We evaluate from the magneto-transport data the temperature dependence of the superconducting critical field, to establish the superconducting phase diagram. 
The  upper critical magnetic field is estimated to be about 35 T at low temperatures from the resistive transition data using the Werthamer-Helfand-Hohenberg formula.
The Hall coefficient $R_{H}$ of  the 48-h-reduced superconducting sample is determined to be -0.5$\times10^{-3}$ cm$^{3}$/C at 30 K, suggesting higher electron concentration. 
These findings have a close relationship with  homogeneous distributions of the superconducting grains and improved weak links between their superconducting grains in the present synthesis process.

\end{abstract}


\renewcommand{\figurename}{Fig.}
\maketitle

\begin{figure}[ht]
\includegraphics[width=9cm]{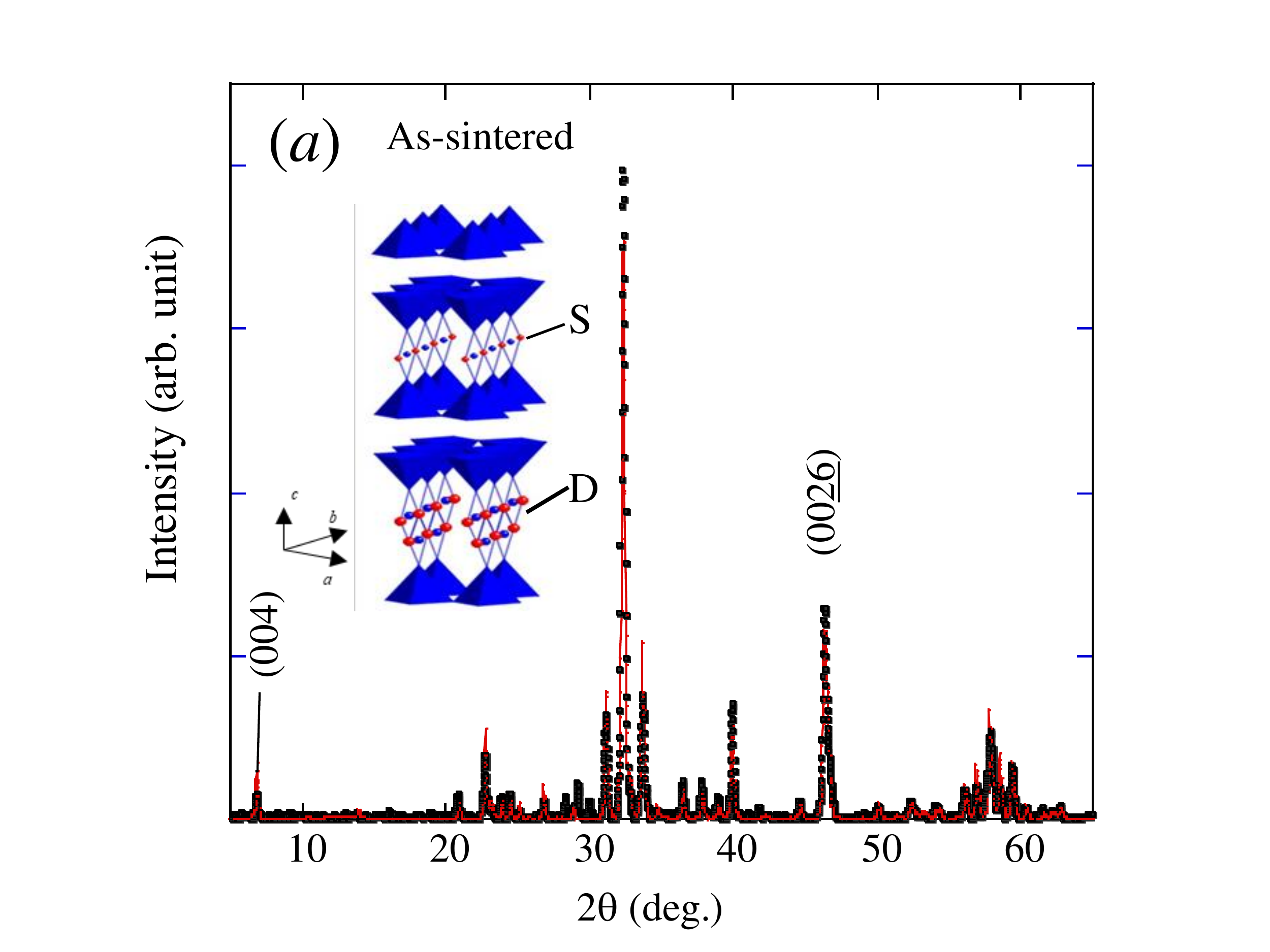}
\includegraphics[width=9cm]{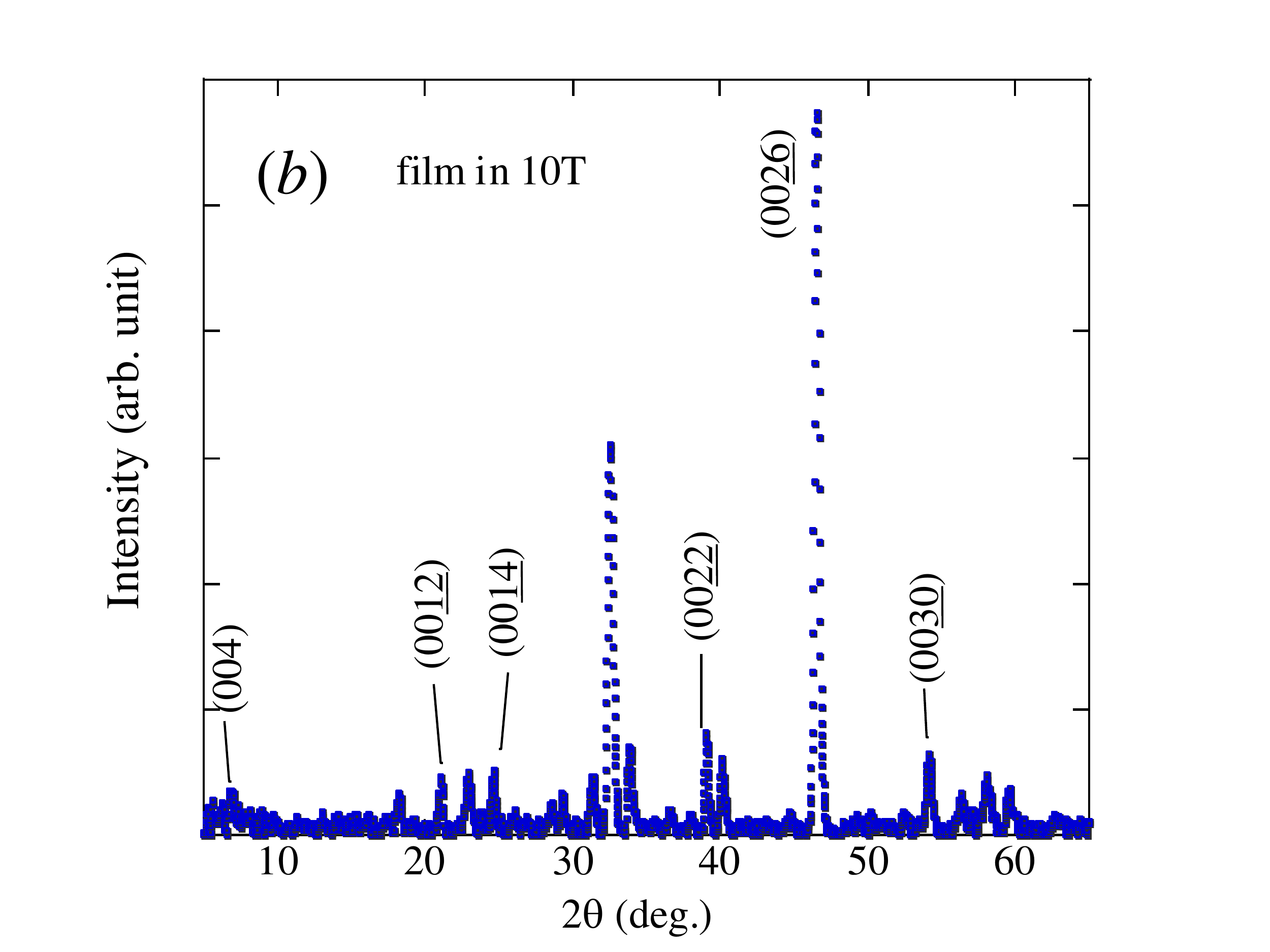}
\caption{(color online)(a) X-ray diffraction patterns of as-sintered polycrystalline  Pr$_{2}$Ba$_{4}$Cu$_{7}$O$_{15-\delta }$. The (004) peak corresponds to one of typical Miller indexes of Pr247.  The calculated curve is obtained using the lattice parameters in the text. Inset shows  the crystal structure of Pr247 with CuO single-chain and double-chain blocks along the $b$-axis. Here, S and D denote CuO single-chain and double-chain blocks along the $b$-axis. (b)X-ray diffraction pattern  for  polycrystalline film prepared by an  electrophoretic deposition technique under 10 T. The magnetic field  was applied   parallel to the direction of applied electric field between anode and cathode electrodes.  }
\label{Xray}
\end{figure}

\begin{figure}[ht]
\includegraphics[width=8cm]{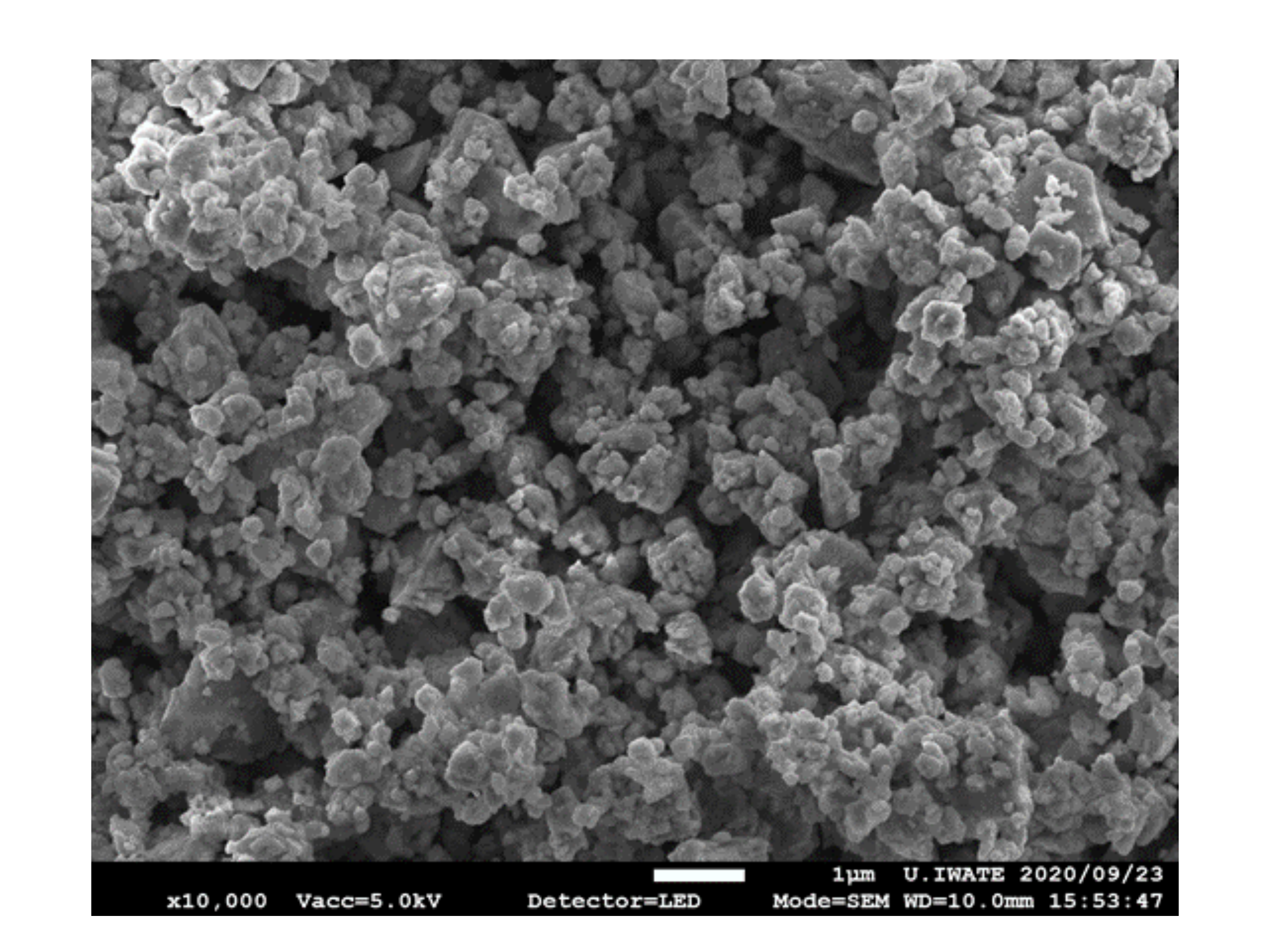}
\caption{(color online) SEM image for  polycrystalline film prepared by an  electrophoretic deposition technique under 10 T. }
\label{SEM}
\end{figure}

\section{INTRODUCTION}
Since the discovery of high-$T_\mathrm{c}$ copper-oxide superconductors,  researches have focused on the unconventional superconductivity 
on the two-dimensional CuO$_{2}$ planes in some cuprates. 
In quasi one-dimensional (1D) ladder system without CuO$_{2}$ planes, it is well known that the superconductivity at  $T_\mathrm{c}$= 12 K  appears only under the application of high pressure.\cite{UE96}
In a previous study of Pr-based cuprates with metallic CuO double chains and insulating CuO$_{2}$ planes
, Pr$_{2}$Ba$_{4}$Cu$_{7}$O$_{15-\delta}$  is found to be a new superconductor  with a higher $T_\mathrm{c}$ (15 K) after a reduction treatment.\cite{MA04}
A nuclear quadrupole resonance (NQR) study has revealed that the newly discovered superconductivity is realized at the CuO double-chain block. \cite{WA05}


Structurally, the Pr-based cuprates, PrBa$_{2}$Cu$_{3}$O$_{7-\delta}$ (Pr123) and PrBa$_{2}$Cu$_{4}$O$_{8}$ (Pr124), are identical to their corresponding
Y-based high-$T_\mathrm{c}$  superconductors, YBa$_{2}$Cu$_{3}$O$_{7-\delta}$ (Y123) and  YBa$_{2}$Cu$_{4}$O$_{8}$ (Y124).
Pr123 and Pr124 compounds have insulating CuO$_{2}$ planes and are non-superconductive. \cite{SO87,HO98}  
The suppression of superconductivity in the Pr substitutes has been explained by the hybridization of 
Pr-4$f$ and O-2$p$ orbitals.\cite{FE93}  
The crystal structure of Pr124 with CuO double chains differs from that of Pr123 with CuO single chains. 
It is well known that CuO single chains in Pr123 and CuO double chains in Pr124 show semiconducting and metallic behaviors, respectively.\cite{MI00}
The carrier concentration of doped double chains of Pr124 is difficult to vary,  because it is thermally stable up to high temperatures.

The compound Pr$_{2}$Ba$_{4}$Cu$_{7}$O$_{15-\delta}$ (Pr247) is an  intermediate between  Pr123 and Pr124.   
In this compound,  CuO single-chain and double-chain blocks are alternately stacked along the $c$-axis such as \{-D-S-D-S-\} sequence \cite{BO88,YA94} (see Fig.\ref{Xray}). 
Here, S and D denote CuO single-chain and double-chain blocks along the $b$-axis, respectively.
The physical properties of  the metallic CuO double chains can be examined 
by controlling the oxygen content along the semiconducting CuO single chains.
Anisotropic resistivity measurements of single-crystal Pr124 have revealed that metallic transport arises by the conduction along the CuO double chains.\cite{HO00}
In oxygen removed polycrystalline Pr$_{2}$Ba$_{4}$Cu$_{7}$O$_{15-\delta }$, the superconductivity appears at an onset temperature $T_\mathrm{c,on}$ of $ \sim $15 K. \cite{MA04} 
Hall coefficient measurements of  superconducting Pr247 with $T_\mathrm{c,on}=15$ K have revealed that at intermediate temperatures below 120 K, the main carriers change from holes to electrons, as the temperature decreases. Accordingly, this compound is an electron-doped superconductor.\cite{MA07} 
In our previous study, we examined the effect of magnetic fields on the superconducting phase of Pr247.\cite{CH13}  Despite of the resistive drop associated with the superconducting transition, we found that the diamagnetic signal was strongly suppressed as expected in  the 1D superconductivity of  CuO double chains. 
We also reported the effect of pressure on magneto-transport properties in the superconducting and normal phases of the  metallic double chain compound Pr$_{2}$Ba$_{4}$Cu$_{7}$O$_{15-\delta }$. \cite{KU16}. The model of slightly warped Fermi surfaces explains not only the magneto-resistance (MR) effect of the non-superconducting sample, but also the pressure-induced MR phenomena of the superconducting sample.


In this paper, we demonstrate the magneto-transport properties of   Pr$_{2}$Ba$_{4}$Cu$_{7}$O$_{15-\delta }$ with  $T_\mathrm{c,on}$ (22-24 K)  accompanied by the strongly metallic conduction, for our understanding of the effect of magnetic field on the superconducting properties of electron-doped metallic double-chain compound.  In the next section, the experimental outline is described. In Sect. 3, we show the outstanding data including higher  residual resistivity ratios, larger superconducting volume fractions, enhanced zero-resistance states under magnetic field, and higher carrier concentrations.  
The final section is devoted to the summary.

\section{EXPERIMENT}
Polycrystalline samples of  Pr$_{2}$Ba$_{4}$Cu$_{7}$O$_{15-\delta}$  were synthesized  by using  the citrate pyrolysis technique.\cite{KO91,HA06} In the first step,  stoichiometric mixtures of high purity Pr$_{6}$O$_{11}$, Ba(NO$_{3}$)$_{2}$,  and  CuO were dissolved in a nitric acid solution at50-60 $^{\circ} $C. After adding  citric acid and neutralizing it by aqueous ammonia,  we then obtained  the porous products  through the self-ignition process using halogen lamp stirrer.  In the next step,   the  precursors  were  ground and resultant fine powders were  annealed  under ambient oxygen pressure at 890-891 $^{\circ} $C for an extended period over 110-120h.
In the present citrate pyrolysis synthesis procedure, we adopted the electric tube furnace with three zone temperature controlled system, to achieve the temperature uniformity within 1 $^{\circ} $C.

For scanning electron microscope (SEM) measurements,  we prepared the polycrystalline  film sample, to reveal microstructures of single-phase  Pr$_{2}$Ba$_{4}$Cu$_{7}$O$_{15}$  powders synthesized by the citrate pyrolysis technique. 
The  polycrystalline film on Ag substrate was fabricated from the as-sintered powders by an  electrophoretic deposition technique. 
The  electrophoretic deposition was conducted in the acetone and iodine bath under the application of  electric voltage up to 300 V for 120 s. \cite{KA01}  We set  Pt and Ag plates as anode and cathode electrodes, respectively. Furthermore,  we performed the magnetic field assisted electrophoretic deposition
process for the fabrication of $c$-axis aligned Pr$_{2}$Ba$_{4}$Cu$_{7}$O$_{15}$ polycrystalline film.\cite{KA02}  Here,  the magnetic anisotropy of rare earth ion Pr including Pr$_{2}$Ba$_{4}$Cu$_{7}$O$_{15}$ shows that  the magnetic susceptibility of  $\chi_{H\parallel c} $ is larger than that of $\chi_{H\perp c} $. \cite{XU10}
A magnetic field up to 10 T was applied to the colloid bath including the suspended Pr247 powders 
using a superconducting magnet with a  100 mm diameter bore at room temperature at the High Field Laboratory for Superconducting Materials, Institute for Materials Research (IMR), Tohoku University.
The magnetic field  was applied parallel to the direction of applied electric field for the electrophoretic deposition. 

We performed X-ray diffraction measurements on the produced samples
at room temperature with an Ultima IV diffractometer (Rigaku) using Cu-K$\alpha $ radiation. 
The lattice parameters were estimated  from the x-ray diffraction data using the least-squares fits.

 
The oxygen in the as-sintered sample was removed by reduction treatment   in a vacuum  at 500 $ ^{ \circ }$C,  yielding  a superconducting material. 
Typical dimensions of the pelletized rectangular sample were $4.2\times 3.4\times 1.3$ mm$^{3}$.
X-ray diffraction data  revealed that the as-sintered polycrystalline samples are an almost single phase with an orthorhombic structure ($Ammm$), as shown in Fig.\ref{Xray}($a$).  

The lattice parameters of the as-sintered sample  are $ a= 3.89293 $ \AA, $ b=3.91394$ \AA, and $ c=50.79270$ \AA,  which are in fair agreement with those obtained by a previous study\cite{YA05}. 
The oxygen deficiencies in the 48 h and 72 h reduced samples in a vacuum  were estimated  from gravimetric analysis to be  $\delta = 0.43$ and 0.54, respectively.
As a function of the oxygen deficiency,  the $T_\mathrm{c,on}$ rises rapidly at $\delta\geq \sim 0.2$, then monotonically increases with increasing $\delta$, and finally saturates  around 26-27 K at  $\delta\geq \sim 0.6$.\cite{HA08}
Accordingly, we expect that the carriers in the present sample are concentrated  around the optimally doped region. 

The electric resistivity in zero magnetic field was measured by the $dc$ four-terminal method. The magneto-transport up to 9 T was measured by the $ac$ four-probe method using a physical property measuring system (PPMS, Quantum Design), increasing the zero-field-cooling (ZFC) temperatures from 4 K  to 40 K. The high field resistivity  (up to 14 T) was measured in a superconducting magnet at IMR, Tohoku University.  The electric current  $I$ was applied longitudinally to the sample ; consequently, 
the applied magnetic field $H$ was transverse to the sample (because $H\perp I$). 
We performed Hall coefficient measurements on the as-sintered and 48 h reduced  samples with the five-probe technique using PPMS  and the 15T-superconducting magnet. \cite{TA13}   
The $dc$ magnetization was performed under ZFC in a commercial superconducting quantum interference  device magnetometer (Quantum Design, MPMS).

\begin{table*}[htb]
\begin{center}
\caption{Physical and superconducting properties  for  Pr$_{2}$Ba$_{4}$Cu$_{7}$O$_{15-\delta }$. 
In details, see the corresponding text and references.  $\delta$ and RRR denote oxygen deficiency and  residual resistivity ratio $\rho$(300K)/ $\rho$(30K), respectively. $f_\mathrm{SC}$: the superconducting volume fraction estimated from the ZFC magnetization data at low temperatures. 
 $H_\mathrm{c}^{*}$ is defined  from the field sweep data at fixed temperatures as the critical field where the zero-resistance state is violated with increasing the field.  
 }

\begin{tabular*}{160mm}{p{25mm}cccccccccccccccc} \hline 
Synthesis&&Reduced time&&$\delta$&&RRR &&$f_\mathrm{SC}$&& $T_\mathrm{c,on}$ &&  $T_\mathrm{c,zero}$ &&$H_\mathrm{c}^{*}$&& $R_{H}$(30K)   \\
&&	&&  && &&  ($\%$) && (K) && (K) &&(T) && ($10^{-3}$ cm$^{3}$/C) \\ \hline
citrate pyrolysis &&48h &&0.43 &&10 &&55 &&22 && 18 &&10.6(4.2K) &&-0.5  \\
citrate pyrolysis&&72h &&0.54&&12 && 50 &&24.3 && 18 &&-&&-  \\ 
citrate pyrolysis&&48h  &&0.52$^{a}$&& 4$^{a}$ && 30$^{a}$ && 26.5$^{a}$ &&15$^{a}$&&$\sim$2(5K)&&-1.1$^{b}$   \\
sol-gel\cite{TO12} &&72h &&0.94&&2 &&$\sim$10  &&30.5 &&$\sim$10 &&-&&-  \\ 
high pressure\cite{MA07} &&48h &&0.5&&- &&30&&16 &&$\sim$10&&- &&-1.5   \\  
\hline 
$^{a}$ see ref. \cite{CH13}, $^{b}$ref.\cite{TA13}   \\

\end{tabular*}
\end{center}
\label{TAB}
\end{table*}

\begin{figure}[ht]
\includegraphics[width=9cm, pagebox=cropbox, clip]{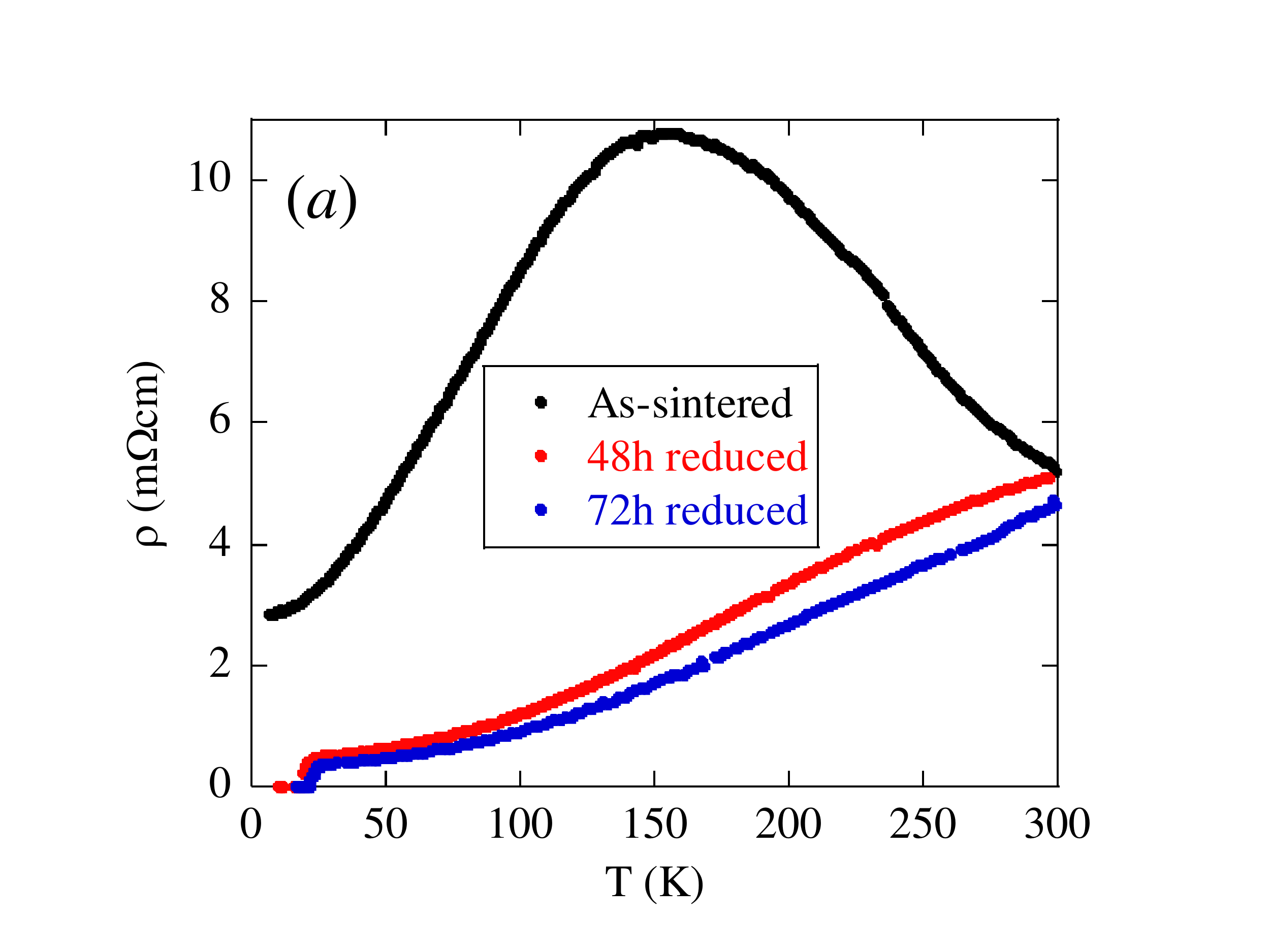}
\includegraphics[width=9cm, pagebox=cropbox, clip]{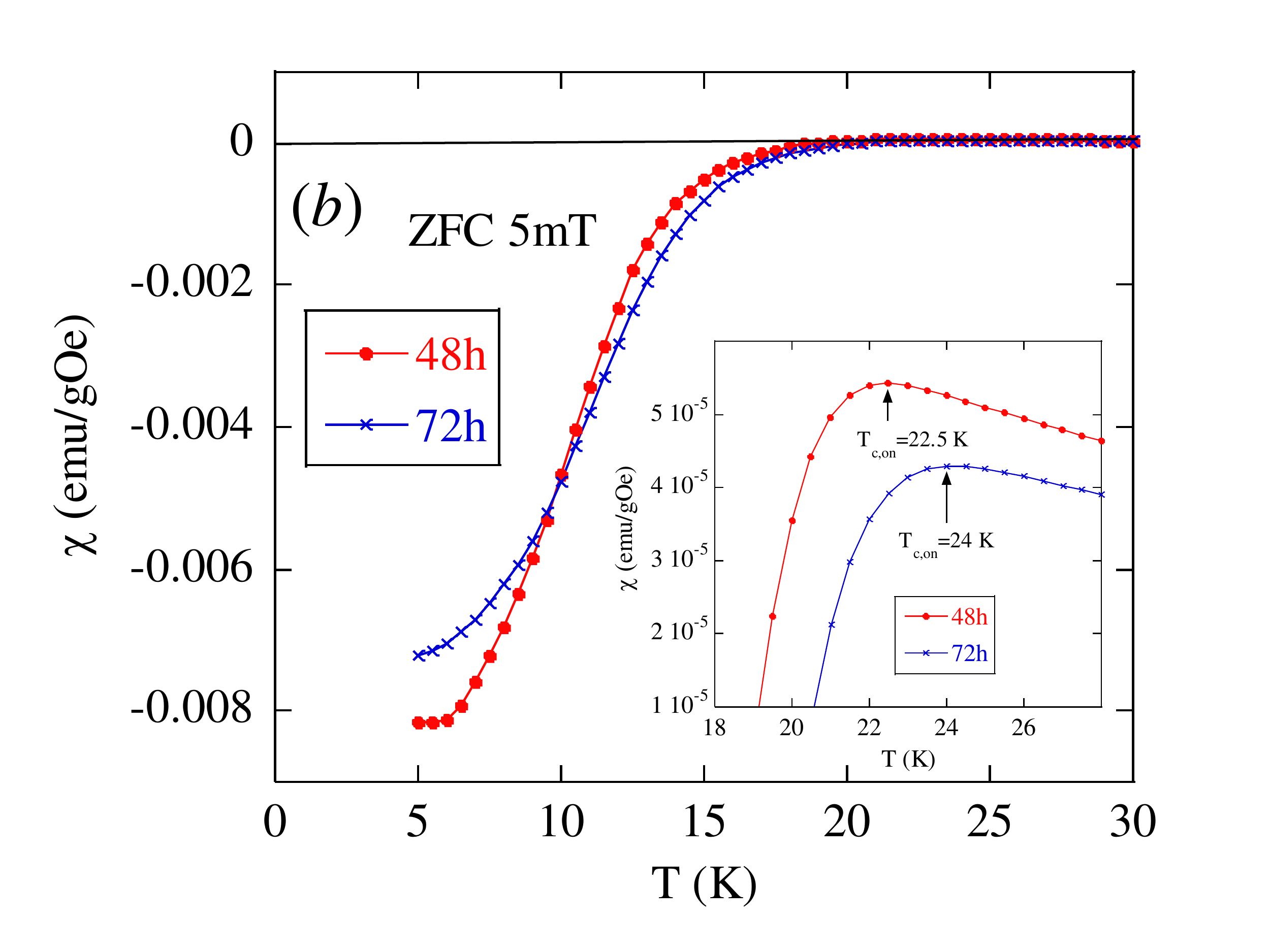}
\caption{(color online)(a) Temperature dependences of electric resistivities of the as-sintered, 48-h-reduced and 72-h-reduced  Pr$_{2}$Ba$_{4}$Cu$_{7}$O$_{15-\delta }$ compounds. 
(b)  low-temperature dependences of magnetic  susceptibilities  $\chi $ of the 48-h-reduced and  72-h-reduced  superconducting samples measured at 5 mT under ZFC scan. 
In the inset, the magnified data are plotted to clarify the definition of $T_{c,on}$.}
\label{RMT}
\end{figure}

\begin{figure}[ht]
\includegraphics[width=9cm, pagebox=cropbox, clip]{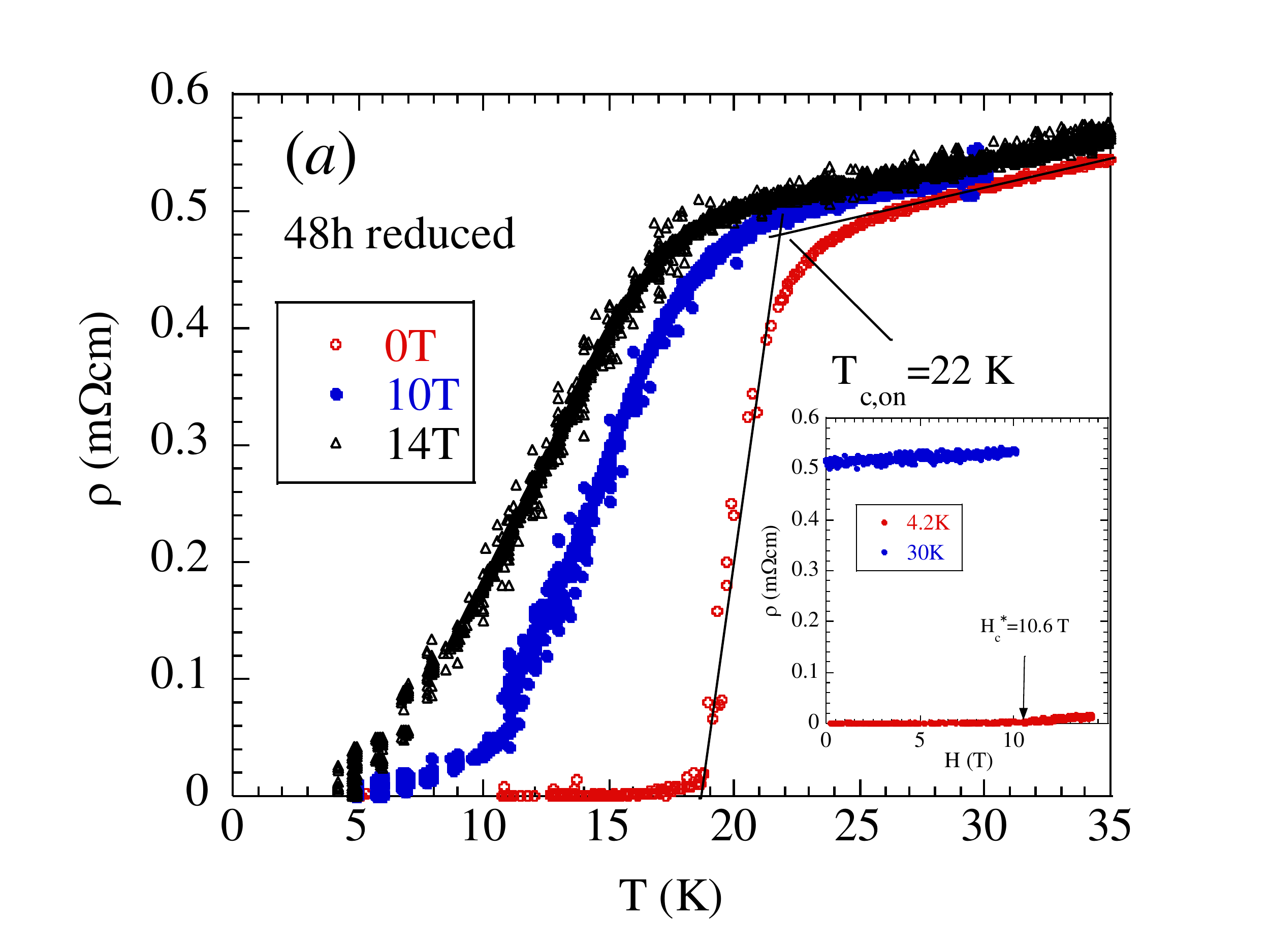}
\includegraphics[width=9cm, pagebox=cropbox, clip]{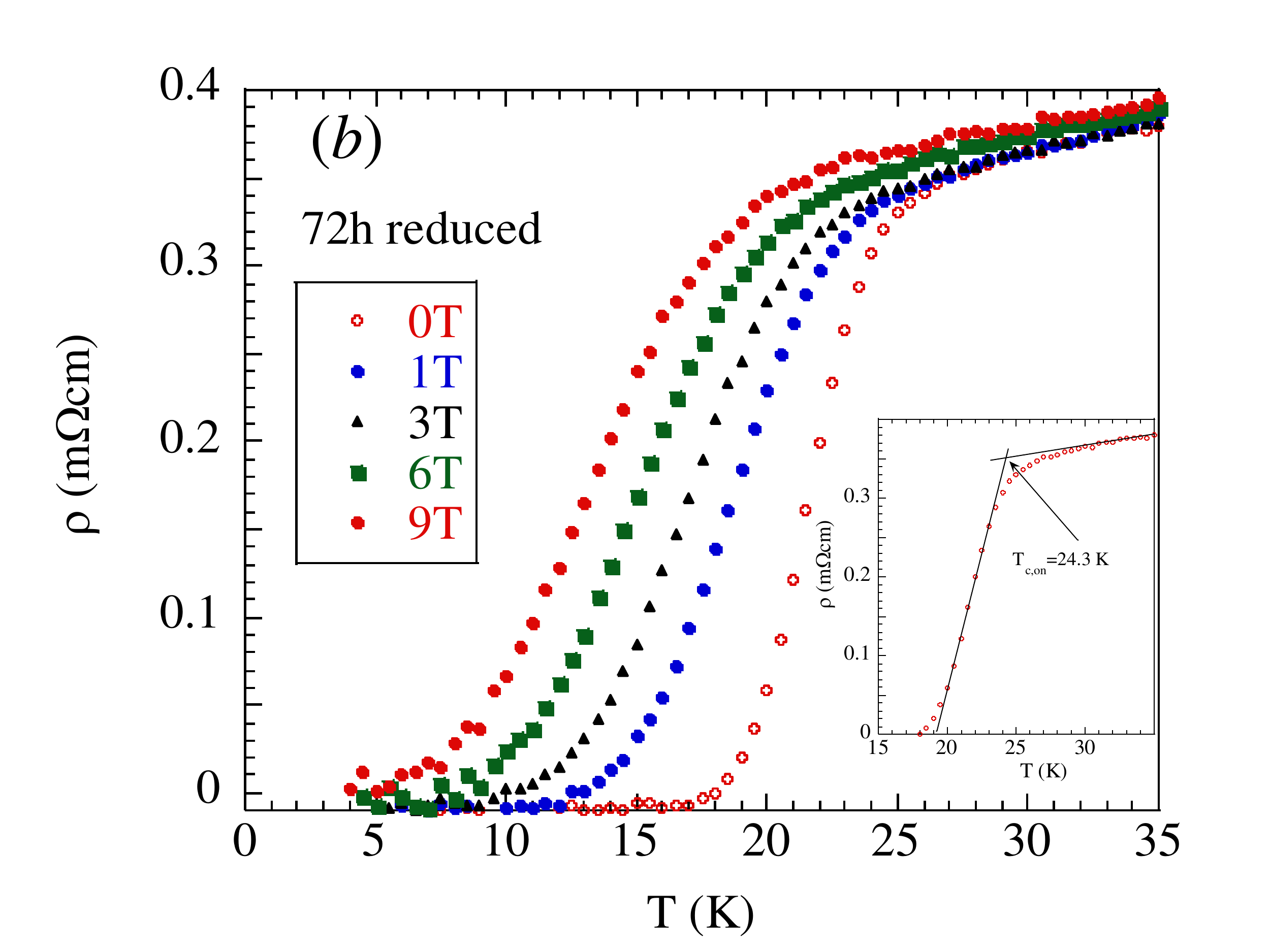}
\caption{(color online)  (a) Low-temperature dependences of electric resistivities of the 48-h-reduced  superconducting Pr$_{2}$Ba$_{4}$Cu$_{7}$O$_{15-\delta }$ compound measured under 0T, 10T, and 14T. Inset plots the magneto-resistance data of the 48-h-reduced sample at 4.2 K and 30 K. 
 (b) Low-temperature dependences of electric resistivities of the 72-h-reduced  superconducting Pr$_{2}$Ba$_{4}$Cu$_{7}$O$_{15-\delta }$ compound measured at the several applied fields ($H$ = 0, 1, 3, 6 and 9 T). The inset displays the enlarged data of the  72-h-reduced sample around $T_{c}$. }
\label{RTH}
\end{figure}

\begin{figure}[ht]
\includegraphics[width=10cm, pagebox=cropbox, clip]{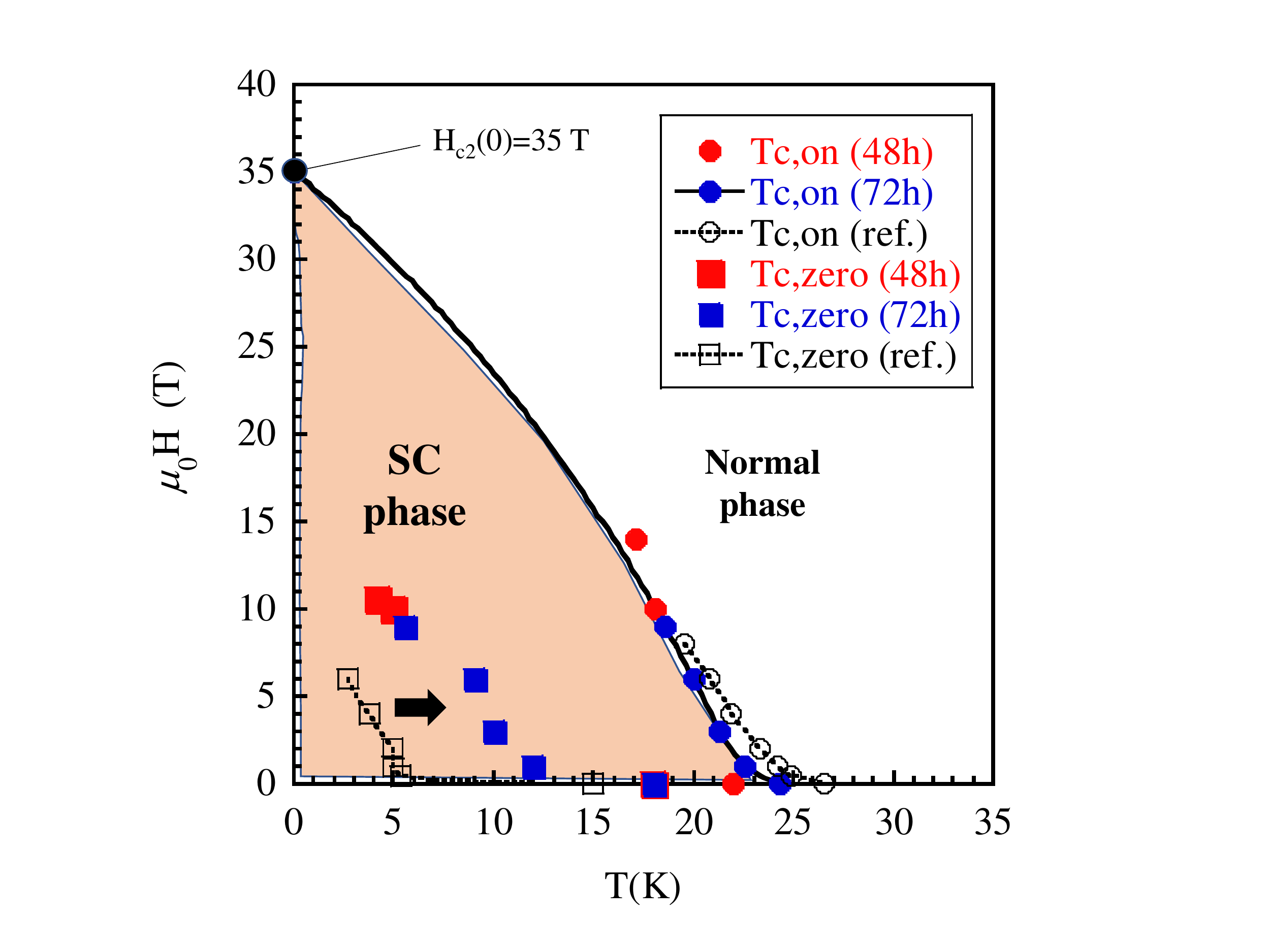}
\caption{(color online)  Temperature dependence of the superconducting (SC) critical field of the reduced samples of   Pr$_{2}$Ba$_{4}$Cu$_{7}$O$_{15-\delta }$. 
The onset $T_\mathrm{c}$  ( $T_\mathrm{c,on}$ )  is determined from the resistivity data as described in the text. For comparison,  the previous data of the reduced Pr247 are referred  (open symbol).
The solid curve separating the superconducting and normal phases was the best fit to the experimental data and the upper critical field of $\sim$35 T.  Here,   the corresponding $H_\mathrm{c2}$(0) is estimated using the WHH formula in the text. }
\label{PHA}
\end{figure}

\begin{figure}[ht]
\includegraphics[width=9cm, pagebox=cropbox, clip]{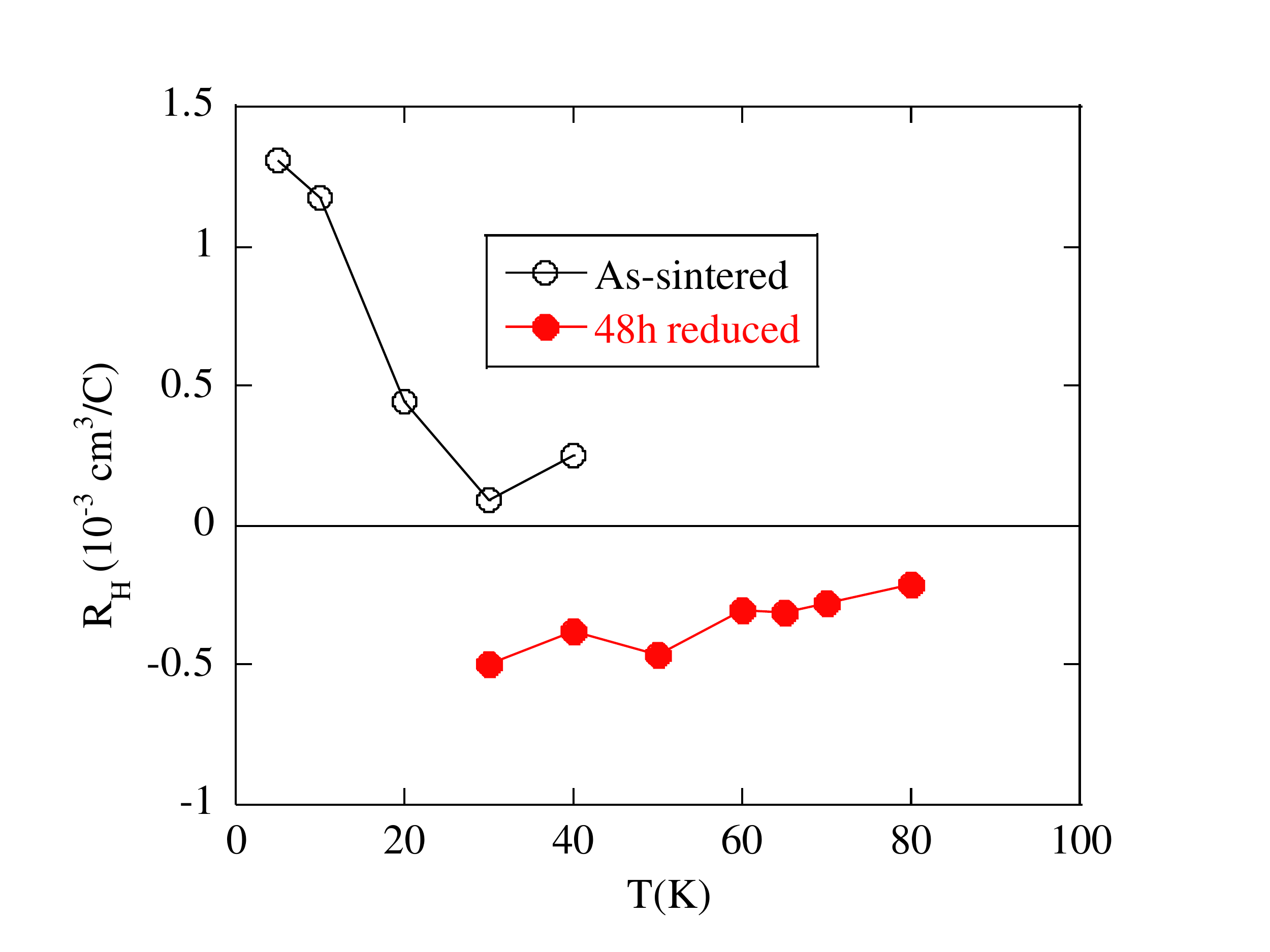}
\caption{(color online) Temperature dependences of the Hall coefficients $R_{H}$ for the as-sintered non-superconducting and 48-h-reduced superconducting samples of  Pr$_{2}$Ba$_{4}$Cu$_{7}$O$_{15-\delta }$.
 }
\label{HAL}
\end{figure}

\section{RESULTS AND DISCUSSION}
Figure\ref{Xray}($a$) shows  X-ray diffraction (XRD) patterns of as-sintered polycrystalline  Pr$_{2}$Ba$_{4}$Cu$_{7}$O$_{15-\delta }$(Pr247). The (004) peak corresponds to one of typical Miller indexes of Pr247.  The calculated curve is obtained using the lattice parameters. The inset of Fig. \ref{Xray}($a$) displays  the crystal structure of Pr247 with CuO single-chain and double-chain blocks along the $b$-axis. 
The unknown impurity peaks near 2$\theta \sim$30  $ ^{ \circ }$ are identified as the BaCuO$_{2}$ phase.\cite{HA06} 
 
For comparison, X-ray diffraction pattern of  polycrystalline film prepared by an  electrophoretic deposition technique under 10 T is shown in  in Fig. \ref{Xray}($b$).  We note that the XRD pattern of the  Pr247 film fabricated by the  electrophoretic deposition process without the applied magnetic field (not shown here)  was the same as that of the Pr247 powders.  (00$l$) peaks did not appear except for the  (004) and (00$\underline{26}$) peaks. 
 When the magnetic field  was applied parallel to the direction of applied electric field,  the peak intensities of  (00$l$)  were strongly  enhanced as shown in Fig. \ref{Xray}($b$), indicating the c-axis alignment of polycrystalline grains.   The SEM image of the Pr247 film in Fig. \ref{SEM} reveals  that  plate-like grains with sub micron size are homogeneously dispersed.

Now,  the temperature dependences of electric resistivities of the as-sintered, 48-h-reduced and 72-h-reduced  Pr$_{2}$Ba$_{4}$Cu$_{7}$O$_{15-\delta }$ compounds are shown in  Fig. \ref{RMT}($a$).
The reduction heat treatment on the as-sintered sample in vacuum results in the appearance of superconducting state with $T_\mathrm{c}$=22-24 K, accompanied by the strongly metallic properties over a wide  range of temperature.
Here, we define the residual resistivity ratio (RRR) as $\rho$(300K)/ $\rho$(30K),  For the present samples reduced in vacuum,  we obtained the higher RRR  values ($10\sim 12.5$), in comparison with the previous data. \cite{CH13} 

Furthermore, to check the bulk superconductivity, we performed to measure low-temperature dependences of magnetic  susceptibilities  $\chi $ of the 48-h and  72-h-reduced  superconducting samples measured at 5 mT under ZFC scan. 
Figure \ref{RMT}($b$) exhibits diamagnetic signals below $T_\mathrm{c,on}$=22.5 K and 24.0 K for the  48-h-reduced and 72-h-reduced samples, respectively.  In addition,  the superconducting volume fractions due to the shielding effect are estimated to be $50\sim55 \%$  from the ZFC values at 5 K, which are much higher than the previous data ($\sim30 \%$). \cite{CH13} 
In the inset of  Fig. \ref{RMT}($b$) , the magnified data are plotted to clarify the definition of $T_\mathrm{c,on}$.  The characteristic values for the present and previous samples are listed in Table I. 

Next,  we try to measure the magneto-transport properties of the reduced samples, to examine the magnetic effect on the superconducting phase of the Pr247 compound. 
 Figure  \ref{RTH}($a$) shows  the low-temperature dependences of electric resistivities of the 48-h-reduced  superconducting Pr$_{2}$Ba$_{4}$Cu$_{7}$O$_{15-\delta }$ compound measured under zero field and applied fields of 10 and 14T. 

In zero field,   the resistivity starts to decrease around $T_\mathrm{c,on}$ = 22 K, then follows a rapid drop, and finally achieves a zero-resistance state at $T_\mathrm{c,zero}$ = $\sim$18 K. 
The observed transition width $T_\mathrm{c,on}$ = 22 K is relatively sharp in comparison with the published data.
Inset plots the magneto-resistance data of the 48-h-reduced sample at 4.2 and 30 K. 
At the high field of 10 T,  we observed $T_\mathrm{c,on}$ = 18 K and  $T_\mathrm{c,zero}$ = $\sim$5 K.
Upon increasing the applied field up to 14 T,  the superconductive critical field at 4.2K is determined to be 10.6 T.  
Even at the maximum field of 14 T, we judge that the zero-resistance state is almost realized over 
the whole sample since the magnitudes of $\rho$ are negligibly small with those of the normal state at 30 K.  These findings strongly suggest that for the present sample  the superconducting properties under  relatively high fields are considerably improved.   For the previous Pr247 samples, it has been reported that the critical field is as low as a few tesla at low temperatures.  

 In Fig. \ref{RTH}($b$) , we show  the low-temperature dependences of electric resistivities of the 72-h-reduced  superconducting Pr$_{2}$Ba$_{4}$Cu$_{7}$O$_{15-\delta }$ compound measured at the several applied fields ($H$ = 0, 1, 3, 6 and 9 T). The inset displays the enlarged data of the  72-h-reduced sample around $T_\mathrm{c,on}=24.3$ K.  In a similar to the data of the 48-h-reduced sample,  we observed a zero-resistance state  below   $T_\mathrm{c,zero}$ = $\sim$18 K  and 5 K,  at  0 T and at 9 T, respectively.
Here, we evaluate from the magneto-transport data the temperature dependence of the superconductive critical field, to establish the superconducting phase diagram. 
The onset $T_\mathrm{c}$  ($T_\mathrm{c,on}$) is determined from the intersection between  the two lines extrapolated from the normal and superconducting transition resistivity data, just above and bellow   $T_\mathrm{c}$, respectively. (see  Fig.\ref{RTH}). The zero-point  $T_\mathrm{c}$  ($T_\mathrm{c,zero}$) denotes the value of the critical temperature reaching the zero-resistance state. 
The upper critical field at 0 K is estimated to be about 35 T from the data of the onset $T_\mathrm{c}$ using the Werthamer-Helfand-Hohenberg (WHH) formula in the dirty limit, $\mu_{0}H_\mathrm{c2}(0)=-0.69T_\mathrm{c}(dH_\mathrm{c2}(T)/dT)|_{T=T_{c}}$.\cite{WE66}

There are quite differences in the critical fields determined from  $T_\mathrm{c,zero}$ 
between the present and previous samples,  strongly suggesting the enhanced superconducting properties in the presence of magnetic field as represented   by a right arrow in Fig. \ref{PHA}.

Furthermore,  in Fig.\ref{HAL}, we show  the temperature dependences of the Hall coefficients $R_{H}$ for the as-sintered non-superconducting and 48-h-reduced superconducting samples of  Pr$_{2}$Ba$_{4}$Cu$_{7}$O$_{15-\delta}$. The $R_{H}$ data of the as-sintered sample  are very similar, in the both  magnitude and temperature dependence, to those of the previous sample. 
For the 48-h-reduced sample,  the  $R_{H}$ data  exhibit  negative values in the limited temperature range between 30 and 80 K,  accompanied by electron doping due to the reduced heat treatment in vacuum.  Moreover,  we estimate  $R_{H}=-0.5\times10^{-3}$ cm$^{3}$/C at 30 K, which is as about half as the published data. \cite{TA13}, indicating higher carrier concentration. 
These results have a close relationship with  the enhanced superconducting properties, leading to 
the superconducting phase diagram of the present samples. 

Finally,  the physical and superconducting properties  for  Pr$_{2}$Ba$_{4}$Cu$_{7}$O$_{15-\delta}$
compounds are summarized in Table I.  
In the present citrate pyrolysis synthesis procedure, we adopted the electric tube furnace with three zone temperature controlled system, to achieve the temperature uniformity within 1 K.  In spite of limited heat-treatment conditions , we obtained homogeneous distributions of the superconducting grains and improved weak links between their superconducting grains, leading to the enhanced superconducting properties in the present samples.  The former is closely related to the narrower transition width of superconductivity in zero magnetic field. The latter improvement is mainly responsible for the considerable enhancement of superconducting properties in high field. 
In addition, we give some comments on the high-$T_\mathrm{c}$  samples synthesized by a sol-gel technique.\cite{TO12} 
The transmission electron microscopy image of the high-$T_\mathrm{c}$ sample with  $T_\mathrm{c,on}$ =30.5 K revealed that there exists an irregular long-period stacking structure along the $c$-axis such as \{-D-D-D-S-S-S-D-D-S-S\} sequences.\cite{CH13}  Here, S and D denote CuO single-chain and double-chain blocks along the $b$-axis. This finding is close to higher concentration, resulting in higher $T_\mathrm{c}$ , in comparison to the regular stacking case of \{-S-D-S-D-S-D\}.  The sol-gel  sample showed  highly inhomogeneous features such as the irregular stacking structures and the coexistence of  the nominal Pr247 phase  with the Pr123 and Pr124 phases. For the corresponding  inhomogeneous sample, we also observed a broad width between  the onset and zero transition temperatures as listed in Table I.

\section{SUMMARY}
We have demonstrated  the enhanced superconducting properties of double-chain based superconductor  Pr$_{2}$Ba$_{4}$Cu$_{7}$O$_{15-\delta}$ synthesized by the  citrate pyrolysis technique using the 3-zone controlled electric furnace.
The outstanding experimental data include the higher  residual resistivity ratios, the larger superconducting volume fractions, the enhanced zero-resistance state under magnetic field, and the higher carrier concentrations. 

The SEM image of the Pr247 film  revealed  that  plate-like grains with sub micron size are homogeneously dispersed. 
For the polycrystalline bulk samples,  we obtained the higher residual resistivity ratio ranging  from  10 to 12 for the 48-72 h reduction. The reduction heat treatment on the as-sintered sample in vacuum results in the appearance of superconducting state with $T_\mathrm{c}$=22-24 K, accompanied by the metallic properties over a wide  range of temperature.
The superconducting volume fractions are estimated from the ZFC data to be 50$\sim55\%$, indicating the bulk superconductivity. 
We evaluated from the magneto-transport data the temperature dependence of the superconductive critical field, to establish the superconducting phase diagram. 
The upper  critical magnetic field was estimated to be about 35 T from the resistive transition data using the WHH formula.
For the 48-h-reduced sample,  the  $R_{H}$ data  exhibit  negative values in the limited temperature range between 30 and 80 K,  accompanied by electron doping due to the reduced heat treatment in vacuum.  Moreover,  we estimated  $R_{H}=-0.5\times10^{-3}$ cm$^{3}$/C at 30 K, which is as about half as the published data, indicative of  higher carrier concentration. 
 In spite of limited heat-treatment conditions for the Pr247 phase , we attained homogeneous distributions of the superconducting grains and improved weak links between their superconducting grains, leading to the enhanced superconducting properties in the present samples.

\section{Acknowledge}
This work was supported in part by MEXT Grands-in-Aid for Scientific Research (JPSJ KAKENHI Grants No. JP19K04995). 
We thank Mr. K. Sasaki for the  SEM measurement and M. Nakamura for his assistance in the PPMS experiments.

\end{document}